\documentclass{llncs} 
\usepackage{makeidx}  
\usepackage{epsfig} 
\usepackage{url}
\usepackage{algorithm}
\usepackage{algorithmic}
\usepackage[latin1]{inputenc} 
\begin{document} 

\mainmatter              
\title{Self-adaptive Gossip Policies for Distributed Population-based Algorithms} 

\author {J.L.J. Laredo\inst{1} \and E.A. Eiben\inst{2} 
\and M. Schoenauer\inst{3} \and P.A. Castillo\inst{1} \and A.M. Mora\inst{1} 
\and F. Fern\'andez\inst{4} \and J.J. Merelo\inst{1}}
\institute{Department of Architecture and Computer Technology \\
University of Granada\\
e-mail: {\tt new@geneura.ugr.es}
\and
Department of Computer Science\\
Vrije Universiteit\\
e-mail: {\tt gusz@cs.vu.nl}
\and
Projet TAO, INRIA Futurs\\
LRI, Univ. Paris-Sud\\
e-mail: {\tt marc@lri.fr}
\and
Department of Computer Science\\
University of Extremadura\\
e-mail: {\tt fcofdez@unex.es} 
}

\maketitle              
\begin{abstract} 
Gossipping has demonstrate to be an efficient mechanism for spreading
information among P2P networks. Within the context of P2P computing,
we propose the so-called Evolvable Agent Model for 
distributed population-based algorithms 
which uses gossipping as communication policy, and represents every
individual as a self-scheduled single thread. The model avoids
obsolete nodes in the 
population by defining a self-adaptive refresh rate which depends on
the latency and   
bandwidth of the network. Such a mechanism balances the migration rate to
the congestion of the links pursuing global population coherence.
We perform an experimental evaluation of this model on a real parallel system 
and observe how solution quality and algorithm speed scale with the number of
processors with this seamless approach. 
\end{abstract} 

\section{Introduction} 

Population-based algorithms are a type of stochastic soft computing 
techniques widely used as problem-independent solvers in typically NP-hard
problems such as graph-search. This paper outlines the general aspects of the
development of distributed population-based algorithms in P2P networks
with an emphasis on  Evolutionary Algorithms (EAs)~\cite{eiben:eas}.

EAs are population-based methods 
with an inherent parallelism that has been widely 
studied (see e.g. 
\cite{cantu:parallelga} for a survey) and falls mainly
under two approaches:  master-slave and the island model. In the master-slave mode the
algorithm runs on the master and the individuals are sent for evaluation
to the slaves, in an approach usually called also {\em farming}. Using the 
island model several independent EAs (islands) 
are used processing their own population, and exchanging the best 
individuals 
between islands with a certain rate \cite{cantu99:topologies}.
Both cases present major adoption problems 
in heterogeneous fully decentralized networks such as P2P networks. 
On one hand, 
master-slave features do not fit with large-scale system robustness 
(master represents a single point of failure)
and scalability (since it depends on evaluation function cost, and has
a bottleneck in the efficiency of the master performing the
evolutionary operations).
On the other hand, P2P systems
do not provide the knowledge of the global environment that the island model
needs  in order to set parameters such as the number of islands, 
the population size per island and the migration rate. Some island
models also need generations in all nodes to run in lockstep, which
calls for homogeneous, synchronized, nodes. This is obviously not the
case in P2P ad-hoc networks. 

Nevertheless, there is a third, finer grained approach, termed Fully 
Distributed 
Model, in which processors host single individuals that evolve on their own. 
Operations that require more than a single individual (e.g., selection
and crossover) take place among a defined set of neighbors (between individuals
on different nodes or available locally to a node) \cite{tomassini:spatially}. 
This model is able to adapt to heterogeneous networks since some P2P 
overlay networks 
\cite{overlaynetworks} provide a dynamic neighborhood whose size grows
logarithmically with respect to the total size of the system in a small-world 
fashion. Following a gossip style, these small-world networks 
spread information in an epidemic manner through the whole 
network (as can be seen in \cite{jelasity:newscast,jelasity:gossip}),
what means that the risk of having obsolete individuals across the
network is minimized as a consequence of the probabilistic global 
``infection'' that the nodes undergo. However, gossiping has to deal with one 
more question: to maintain the larger coherence in a distributed population 
among the network, which implies locally to a node not only having high 
probability of being ``infected'' but also frequently ``infected''. 
We present the approach of the Evolvable Agent model for dealing with such 
questions.

It is obviously not straightforward to outline a method that takes
advantage of those P2P properties, obtaining at the same time high
performance and good scalability. 
That is why we propose a self-adaptive
refresh rate over the basis of a gossip scheme which balances the frequency 
of ``infections'' to the congestion 
of links. Within this model, each individual in an evolutionary computation 
population rises to an autonomous agent by scheduling its own
actions.

The main objective of this paper is to provide an empirical assessment of our 
agent-based evolutionary model which is a step towards
a ``Fully Distributed Model'' for designing EAs in heterogeneous networks.
To this end, we perform an experimental evaluation on 
a real parallel scenario with up to 6 nodes. 

The rest of the paper is organized as follows: next (section
\ref{sec:work}) we describe the
state of the art in P2P evolutionary computation and other related
subjects. The model is described in section \ref{sec:model}, and the
particulars of the evolutionary algorithm used here are described in
\ref{sec:tournament}. Finally, experimental setup is presented in
section \ref{sec:setup}, results in section
\ref{sec:result} and some conclusions drawn in \ref{sec:conclusions}.

\section{Related Work}
\label{sec:work}

Due to the diversity of fields that this study involves, it is convenient to 
revise them in order to set the scope of the work.

Concerning development of P2P distributed computing systems, there are some 
frameworks such as:
\begin{itemize}
\item DREAM \cite{arenas:dream}, which focuses in distributed processing of 
EAs and uses the P2P network DRM.
\item G2DGA \cite{g2dga}, equivalent to the previous. It centers on 
distributed genetic 
algorithms processing by the use of the network G2P2P.
\item JADE (Java Agent Development Framework, available from
\url{http://jade.cselt.it/}), a P2P system which includes agents as
software components.  
\end{itemize}

The mentioned DRM is an implementation of the newscast protocol 
\cite{jelasity:newscast}. This 
protocol has served as a guide for the proposed communication mechanism within 
this work. 
Newscast is an epidemic
 approach where every node shares local information with its neighbourhood
by selecting a node from it with uniform probability each certain time 
(refresh rate).
Our communication model is inspired by such a protocol. However, 
our model considers a dynamic refresh rate which depends on the QoS parameters:
latency and bandwidth.

Related to agent-based systems for evolutionary computation, 
Vacher et al. present in \cite{multiagent} a multiagent approach to solve 
multiobjective problems. It also describes the implementation of 
functions and operators of the system.
There are some works regarding optimization of parallel evolutionary 
algorithms; Viveros and Bar\'an \cite{agcombinados} propose the combination of 
parallel evolutionary algorithms with local optimization functions which 
depends on processor capacities in heterogeneous computational systems. 
The authors have published related papers on this field: \cite{Fernandez:balancing} 
shows that the 
number of parallel 
executions must be equivalent to the number of available processors in order 
to equilibrate computational effort and algorithmic results. \cite{eiben:onthefly} 
report the benefits of considering population size 
adjustment on runtime. Finally, \cite{laredo:autoadaptacion}
 presents a model (also an agent-based system) where the load
 of every evolutionary computation experiment is self-adaptive depending 
on the architecture where it is executed, yielding more efficient results 
than the classical sequential approach.

In this paper we study an agent-based approach for distributed evolutionary
algorithms and propose an 
asynchronous communication method that allows self-adaptation to 
different network scenarios and dynamic environments such as P2P systems.

\section{Overall Model Description}
\label{sec:model}

\begin{figure} 
\begin{center} 
\epsfig{file=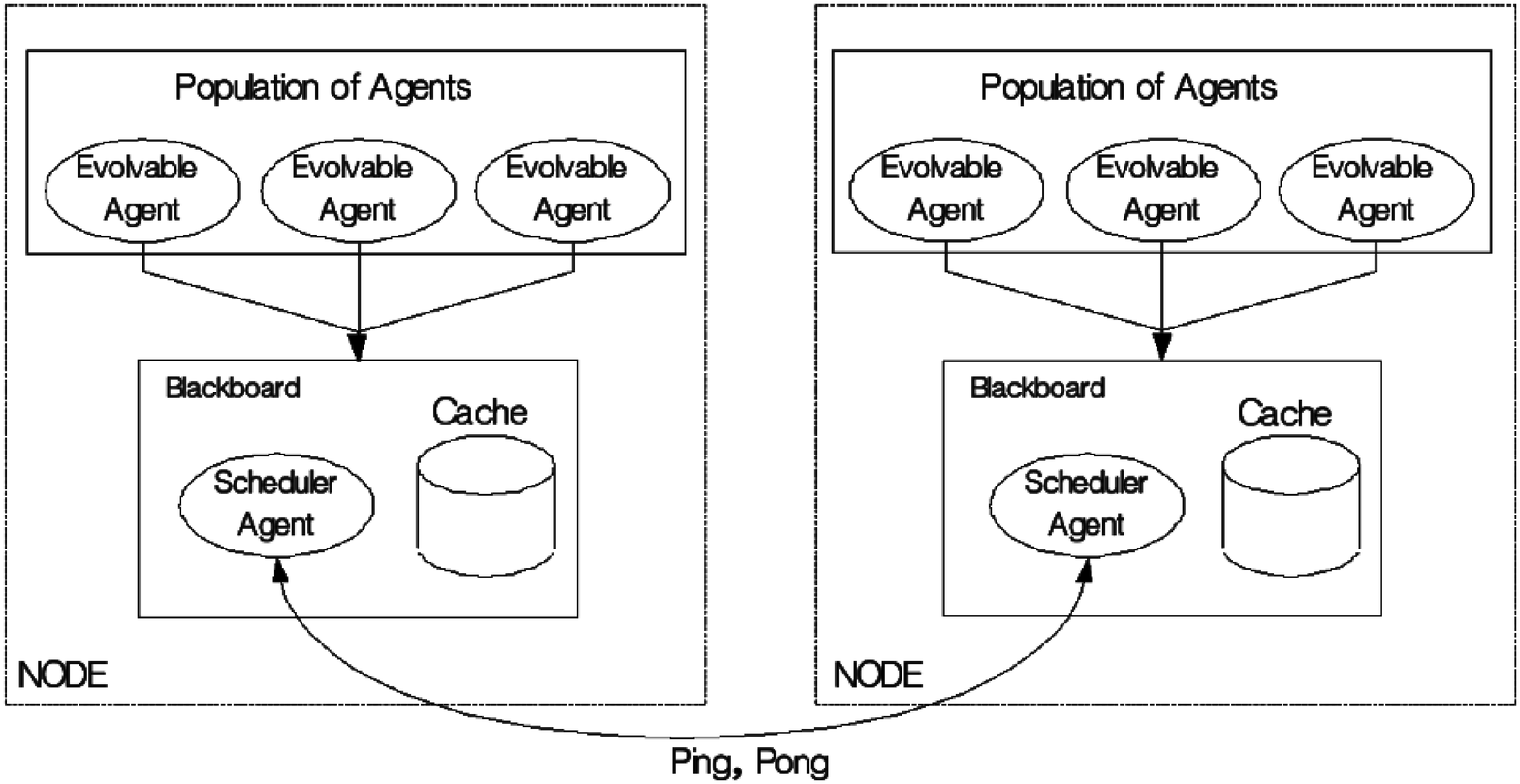,width=8cm}
\end{center}
\caption{Overall architecture of the model \label{fig:architecture}} 
\end{figure}

\begin{figure} 
\begin{center} 
\epsfig{file=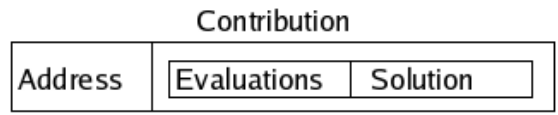,width=8cm}
\end{center}
\caption{Format of a cache entry. It provides the following information about 
a foreign node: Address, number of evaluations performed and one 
individual of its population termed solution
\label{fig:contribution}} 
\end{figure}

The overall architecture of our Evolvable Agent Model is depicted on
figure \ref{fig:architecture}.  
It consists of a group of Evolvable Agents (each one running on its
own thread) whose main design objective is to carry out the principal steps of 
evolutionary computation: selection and variation (crossover and mutation). 
Obviously, the key element here is the locally executable selection. Crossover and
mutation never involve many individuals, but selection in EAs usually requires 
a comparison among all individuals in the population. Consider, for example, roulette wheel or
rank-based selection.

The agents know the environmental status by means of a blackboard
mechanism \cite{blackboard}. The blackboard allows the interchange of
information between agents (Agent-Agent) or with cache
(Agent-Cache). Furthermore, the blackboard implements a Scheduler Agent
that allows information spread among nodes in a gossip style.
The
messages used among nodes are called {\em contributions} and their structure
matches with a cache entry (figure \ref{fig:contribution}).  Thus,
instead of the classical view of a population of solutions managed by an 
oracle, this model proposes a population of agents, each one representing a 
solution.

\subsection{Self-Adaptive Gossip Mechanism}

\begin{algorithm}
\caption{Scheduler Agent}
\label{alg1}
\begin{algorithmic}

\STATE $\Delta T \Leftarrow$ 1 sec.
\LOOP
\STATE sleep $\Delta T$
\STATE Node $\Leftarrow$ Random selected node
\STATE Sol $\Leftarrow$ Selected Solution in $P_{agents}$
\STATE Contribution $\Leftarrow$ $Num_{Evaluations}$,Sol
\STATE Ping (Node,Contribution)
\ENDLOOP
\end{algorithmic}
\end{algorithm}

\begin{algorithm}
\caption{Ping Handler}
\label{alg2}
\begin{algorithmic}

\REQUIRE Node, Contribution
\STATE Cache(Node)$\Leftarrow$ Contribution
\STATE Pong(Node,OK)

\end{algorithmic}
\end{algorithm}
\begin{algorithm}
\caption{Pong Handler}
\label{alg3}
\begin{algorithmic}

\REQUIRE Node
\STATE $\Delta T \Leftarrow$ Time used answering the Ping

\end{algorithmic}
\end{algorithm}

Algorithms \ref{alg1},\ref{alg2} and \ref{alg3} show the pseudo-code of the 
main tasks in the 
communication process. Each blackboard 
maintains a cache with a maximum of one entry per
node in the network. Each entry follows the contribution format (Figure 
\ref{fig:contribution}). The cache indexes the entries with the Address field.
Therefore, the newest contributions replaces the oldest ones. This process 
leads the removal of obsolete individuals and allows a global evolution in
a decentralized environment.
 The scheduling mechanism is carried out by each node as explained next:

\begin{itemize}
\item \textbf{Algorithm \ref{alg1}} Each $\Delta T$ time, a node (the current 
node) selects another node with uniform probability to establish 
communication. Current node sends an application level {\sf Ping} message to 
the selected node with information about a random solution in the population of
 agents ($P_{agents}$) in a contribution format (Figure \ref{fig:contribution}).
\item \textbf{Algorithm \ref{alg2}} The selected node stores that solution
in its cache and sends back an acknowledge message ({\sf Pong}).
\item \textbf{Algorithm \ref{alg3}} At the arrival of the Pong, the current node 
updates its refresh rate ($\Delta T$) with the time spent in the operation.
\end{itemize}

\subsection{Evolvable Agent with Tournament Selection}
\label{sec:tournament}

\begin{algorithm}
\caption{Evolvable Agent with Tournament Selection}
\label{alg4}
\begin{algorithmic}

\STATE $S_{t}$ $\Leftarrow$ Initialize Agent
\STATE Register Agent on the blackboard

\LOOP
\STATE Sols  $\Leftarrow$ Selection($k$, Blackboard)
\STATE $S_{t+1}$ $\Leftarrow$ Recombine(Sols,$P_{c}$) 
\STATE $S_{t+1}$ $\Leftarrow$ Mutate($S_{t+1}$, $P_{m}$)
\STATE $S_{t+1}$ $\Leftarrow$ Evaluate($S_{t+1}$) 
\IF{$S_{t+1}$ better than Blackboard.BestSol}
\STATE    Blackboard.BestSol $\Leftarrow$ $S_{t+1}$ 
\ENDIF
\IF{ $S_{t+1}$ better than $S_{t}$} 
\STATE    $S_{t}$ $\Leftarrow$ $S_{t+1}$
\ENDIF
\ENDLOOP
\end{algorithmic}
\end{algorithm}

Algorithm \ref{alg4} shows the pseudo-code of an Evolvable Agent which uses
Tournament Selection.
The agent owns a solution ($S_{t}$) which it tries to evolve. 
The selection mechanism works as follows:
Each agent selects $k$ ($k$ = tournament size) solutions 
among other agents' 
current solutions and solutions stored in cache (which are migrants from 
network nodes) with uniform probability by means of the blackboard.
The two best solutions are stored in ``Sols'' ready to be recombined by
a crossover operator. The crossover returns a single solution $S_{t+1}$ that
is mutated and evaluated. If the newly generated solution $S_{t+1}$ is better than
the old one $S_t$, it becomes the current solution. 
Finally, Blackboard maintains global elitism by storing the best solution found so far
in Blackboard.BestSol.

\section{Experimental Setup}
\label{sec:setup}


We have carried out an empirical investigation over the Evolvable Agent Model
through conducting experiments on a real parallel scenario with up to 6 node.

As a test problem we have chosen the Travelling Salesman Problem (TSP) 
\cite{tsp}.
The TSP is a classical combinatorial 
optimization 
problem widely used to test evolutionary algorithms 
\cite{wang98:comparative}. In this problem there is a set of $N = 1,\dots,n$ 
cities which have to be visited once in such a manner that the path forms a 
graph cycle that minimizes the travelled distance. We have selected
three symmetrical instances with different complexities: \emph{bier127},
\emph{d198} and \emph{lin318}, extracted from TSPLIB\footnote{http://www.iwr.uni-heidelberg.de/groups/comopt/software /TSPLIB95/ Accessed on January 2007}.


\begin{table}
\centering
\begin{tabular}{|c|c|}
\hline
\texttt{Test-bed}&\\
\hline
Number of Nodes&6\\
Node Processor& 2X AMD Athlon(tm)2400+\\
Network& Gigabit Ethernet\\
Operating System & Linux 2.6.16-1.2096\_FC5smp\\
Java version & J2RE (build 1.5.0\_06-b05)\\
\hline
\hline
\hline
\texttt{Evolutionary Algorithm}&\\
\hline
Crossover & Order crossover(OX)\\
Mutation & 2-Opt mutation\\
Termination condition & Max. number of eval.\\
\hline
\hline
\texttt{Parameter values}&\\
\hline
Population size& 32\\
Tournament size& 7\\
Number of evaluations& 1000000\\
Probability of crossover& 0.7\\
Probability of mutation & 0.1\\
\hline
\end{tabular}
\caption{Test-bed and Evolutionary Algorithm settings.
\label{table:parallel}}
\end{table}

This experiment will provide  data on how solution 
quality (accuracy) 
and algorithm speed scale with the number of processors. Therefore, we 
compare results obtained on a single node up to 6 nodes. (Trivial 
practicalities hindered testing larger networks; further scale-up
test are being prepared.) The physical test-bed and the EA main features for 
the parallel scenario are shown in Table \ref{table:parallel}. 
Solution quality is measured by the mean best fitness (MBF) over 30 
independent runs.
We calculate the speed-up as $S_n=\frac{T_1}{T_n}$, where $T_i$ is the time in 
seconds
spent to reach the termination condition when using $i$ nodes.
Linear speed-up is a reference and we use it as the baseline for comparing 
the scalability results.

Due to the small number of available nodes ($n = 1, \dots, 6$), we used
for all experiments fully connected graph topologies instead of P2P overlay
networks. Such scenario grows with a complexity $O(\frac{(n-1)n}{2})$ which
 intensifies the impact of communication overhead since a real 
P2P 
overlay network should grow with a smaller order of complexity in a
small-world fashion. As designed, the Scheduler Agent will self-adapt the
refresh rate $\Delta T$ to the congestion of links.

\section{Experimental Results}
\label{sec:result}

This experimental evaluation focuses on the analysis of the Evolvable Agent 
model when it scales up to 6 nodes.

Related literature (\cite{Fernandez:balancing} i.e.) shows how algorithmic
results differ in distributed EAs depending either on the availability of
computing resources and the spread of the population. A t-Student 
analysis over the best fitness distributions on the three instances under 
study reveals that there are no significant differences between most of the 
them:
\begin{itemize}
\item In the {\sf bier127} instance, the fitness distributions in 3, 4, 5 and 6
nodes  do not present significant differences.
\item The same happens in {\sf d198} in the case of 1 and 2 
nodes, and 4 and 5 nodes.
\item In {\sf lin318}, the fitness distributions in 2, 4, 5 and 6 nodes 
do not show  significant differences.
\end{itemize}

Since distributed EAs suffer structural changes at population level which
modify their algorithmic behaviour, we can conclude from the previous 
observations that the model under study minimizes the impact of having a
distributed population by means of the Scheduler Agent. We have to take into 
account that the test-bed is composed of a high availability network and 
the Scheduler Agent adapts migration rates to the network latency and
bandwidth. It should be taken into account, too, that differences could be
mostly due to the effects of a small population in the algorithm
result and might be fixed by using a larger population.

\begin{figure} 
\begin{center} 
\epsfig{file=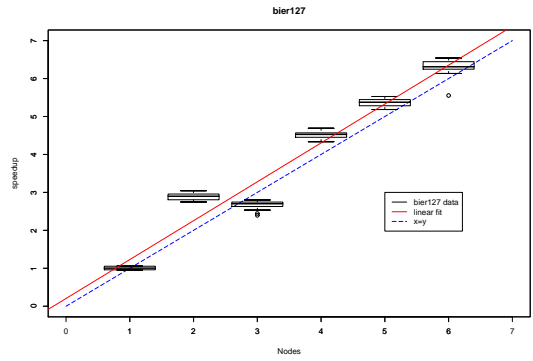,width=5cm,angle=-90}
\end{center}
\caption{Speed-up of the Model up to 6 nodes vs. linear speed-up.
TSP instance {\sf bier127}.
\label{fig:bier127}} 
\end{figure}

\begin{figure} 
\begin{center} 
\epsfig{file=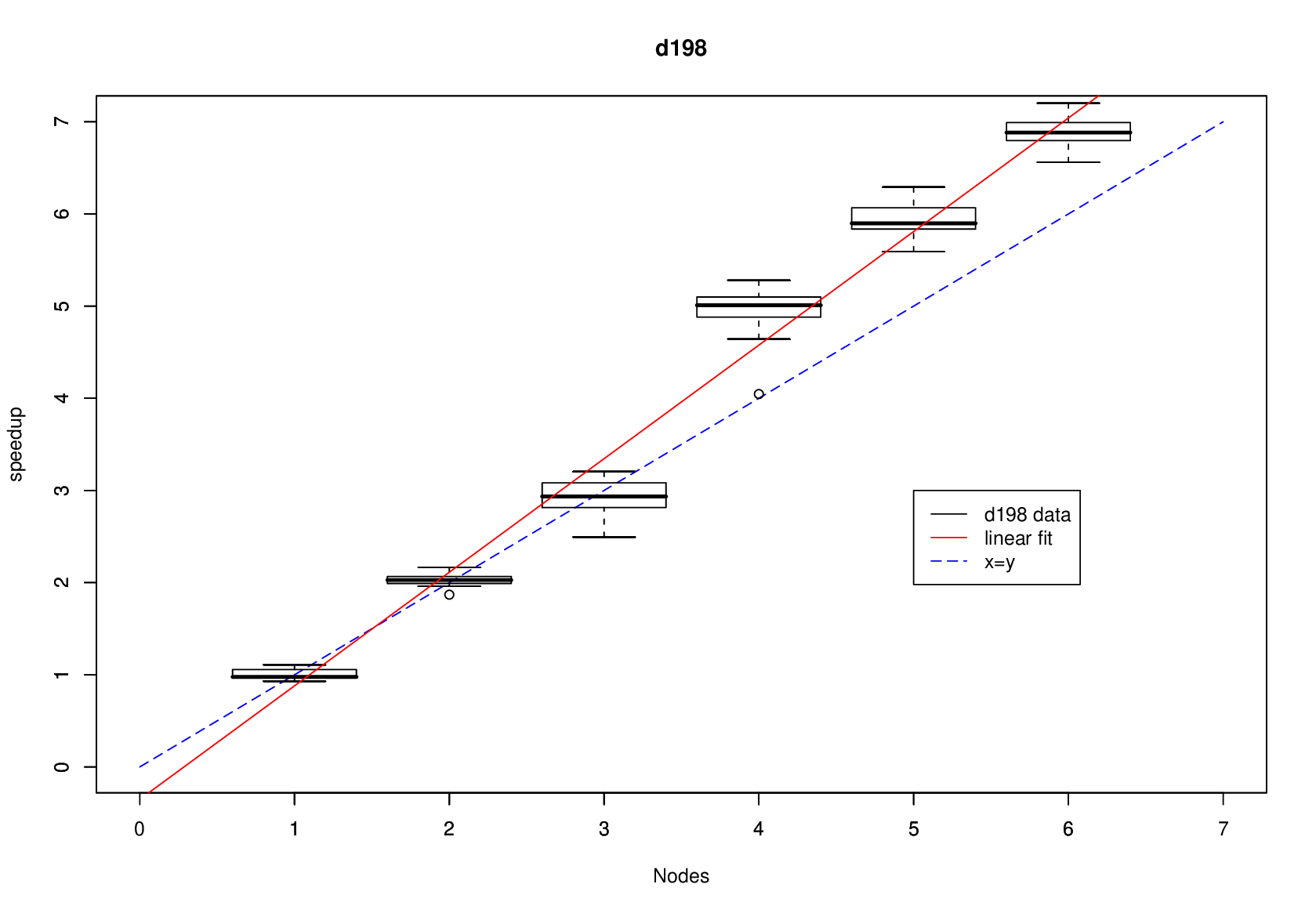,width=5cm,angle=-90}
\end{center}
\caption{Speed-up of the model up to 6 nodes vs. linear speed-up.
TSP instance {\sf d198}.
\label{fig:d198}} 
\end{figure}

\begin{figure} 
\begin{center} 
\epsfig{file=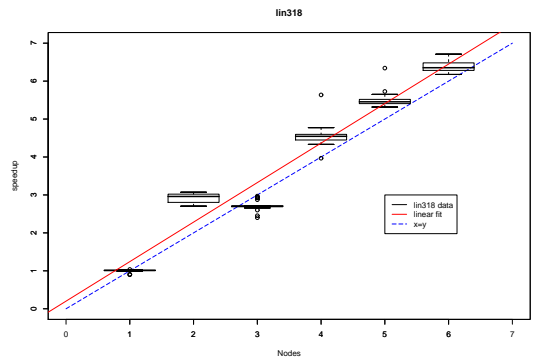,width=5cm,angle=-90}
\end{center}
\caption{Speed-up of the model up to 6 nodes vs. linear speed-up.
TSP instance {\sf lin318}.
\label{fig:lin318}} 
\end{figure}

Concerning scalability, figures \ref{fig:bier127}, \ref{fig:d198} and 
\ref{fig:lin318} represent the speed-up for all problem instances. 
The linear fit over the data shows a growth close to the baseline 
(but a bit over it). The data show that the algorithm speed scales well, 
while maintaining solution
quality (as has been previously shown). Unfortunately, 6 is still a
small network, for extrapolating these results more research is
needed. However, within the bounds of this experiment, we consider
proved that the Autonomous Agent Model has an efficient scaling
behavior, adapting seamlessly to a concurrent as well as a distributed
environment.


\section{Conclusions and Future Work}
\label{sec:conclusions}

In this paper we present an Agent-based approach towards a fully 
distributed EA model. The model is designed to deal with heterogeneous networks
features
and specially P2P networks. The 
evolution process consists in maintaining a population of agents that evolve 
single solutions. 
Each agent can access other agents' current solution in operations that needs 
more than one individual (e.g. selection) by means of the blackboard mechanism 
described in section \ref{sec:model}.

From the proposed experiments we conclude that the model scales with linear 
gain up to 6 nodes despite the growing complexity in topologies. For a 
pre-established computational effort, the best 
fitness distributions in the different test topologies do not reveal 
significant differences in most of the cases. Therefore, the Evolvable Agent
model is a distributed EA model where
scalability and quality results are possible both together. Furthermore,
such an approach is worth as a proof of concept about self-adaptive gossiping
policies for establishing asynchronous migration rates in population-based 
algorithms.


Future works will have to consider the experimentation in large-scale networks
where further conclusions can be reached respecting scalability limitations, 
adaptation to heterogeneity and algorithmic effects of having high latency 
links. Within this line we plan to implement the model 
into a P2P framework such as DREAM \cite{arenas:dream} which shares
its main 
design objectives.


\section *{Acknowledgements}
This work has been partially supported under the Project NadeWeb 
(TIC2003-09481-C04).

\bibliographystyle{plain}
\bibliography{europar2007} 
 
\end{document}